\documentclass[aps,prb,preprintnumbers,amsmath,amssymb,superscriptaddress,twocolumn,10pt]{revtex4-2}

\usepackage{txfonts}
\usepackage{graphicx}
\usepackage{dcolumn}
\usepackage{bm}
\usepackage{array}
\usepackage[dvipsnames]{xcolor}
\usepackage[%
    colorlinks=true,
    pdfborder={0 0 0},
    linkcolor=blue,
    citecolor=black
]{hyperref}  
\usepackage{chngpage}
\usepackage{appendix}

\usepackage{blindtext,subcaption,graphicx}
\usepackage{booktabs}
\usepackage{caption}
\usepackage[T1]{fontenc}
\usepackage{lmodern}

\definecolor{Dark}{gray}{0.2}
\definecolor{MedDark}{gray}{0.4}
\definecolor{Medium}{gray}{0.6}
\definecolor{Light}{gray}{0.8}
\definecolor{darkred}{rgb}{0.55, 0.0, 0.0}
\definecolor{darkslateblue}{rgb}{0.28, 0.24, 0.55}
\definecolor{royalblue(web)}{rgb}{0.25, 0.41, 0.88}
\usepackage{graphicx,float}
\usepackage{amsmath}

\usepackage{amssymb}
\hypersetup{
    citecolor=darkred,
    linkcolor=blue,   
    urlcolor=blue}

\usepackage{bbm}

\def\be{\begin{equation}}
\def\ee{\end{equation}}
\def\beq{\begin{equation}}
\def\eeq{\end{equation}}
\def\bea{\begin{eqnarray}}
\def\eea{\end{eqnarray}}
\def\nbea{\begin{eqnarray*}}
\def\neea{\nonumber\end{eqnarray*}}
\def\bmat#1{\left(\begin{array}{#1}}
\def\emat{\end{array}\right)}
\def\bcase#1{\left\{\begin{array}{#1}}
\def\ecase{\end{array}\right.}
\def\bmini#1{\begin{minipage}{#1\textwidth}}
\def\emini{\end{minipage}}

\usepackage{wrapfig}
\graphicspath{{figures/}}
\usepackage{enumerate}

\begin{document}
\raggedbottom
\setlength{\abovedisplayskip}{5pt}
\setlength{\belowdisplayskip}{5pt}
\setlength{\abovedisplayshortskip}{0pt}
\setlength{\belowdisplayshortskip}{0pt}

\title{Evolution from Kitaev to XXZ spin chains via distortion: Application to BaCo$_2$V$_2$O$_8$}
\author{Philip Richard}
\affiliation{Department of Physics, University of Toronto, Ontario, Canada M5S 1A7}
\author{Mandev Bhullar}
\affiliation{Department of Physics, University of Toronto, Ontario, Canada M5S 1A7}
\author{Hae-Young Kee}
\affiliation{Department of Physics, University of Toronto, Ontario, Canada M5S 1A7}
\affiliation{Canadian Institute for Advanced Research, CIFAR Program in Quantum Materials, Toronto, Ontario, Canada, M5G 1M1}
\date{\today}

\begin{abstract}
BaCo$_2$V$_2$O$_8$ is a prototypical spin-orbit-coupled Ising-chain antiferromagnet that provides a unique platform for studying field-induced quantum magnetism. Under a transverse magnetic field, it exhibits unusual magnetic properties, including an anomalous staggered magnetization and a pronounced anisotropy of the critical field with respect to the in-plane field direction. While these phenomena have been attributed phenomenologically to a site-dependent anisotropic $g$-tensor, a recent microscopic theory has shown that spin-orbit coupling naturally generates bond-dependent Heisenberg, Kitaev, and $\Gamma$ exchange interactions. Here, we unify these two pictures by extending the microscopic theory to incorporate distortions of the CoO$_6$ octahedra. We show that the distortions not only generate the site-dependent anisotropic $g$-tensor but also renormalize the staggered exchange interactions through a distortion-induced contribution that partially compensates the Kitaev-derived staggered term. Using point-charge calculations to estimate the $g$-tensor of BaCo$_2$V$_2$O$_8$, we demonstrate that the strong anisotropy of the critical field originates from the combined effects of the modified exchange interactions and the anisotropic $g$-tensor. Our work provides a unified microscopic framework for understanding the magnetic anisotropy of spin-orbit-coupled Ising-chain materials.
\end{abstract}
\maketitle

\section{Introduction}
The one-dimensional (1D) Ising model has long been a prototypical example in the study of magnetism as well as critical phenomena. The classical case was solved over a hundred years ago and was shown to have no phase transition \cite{Lenz1920, Ising1925}. On the other hand its quantum counterpart, solved exactly via a Jordan-Wigner transformation, undergoes a quantum phase transition in the presence of a transverse magnetic field and possesses fractionalized domain-wall excitations in the ordered phase \cite{Pfeuty1970AoP, henkel1995PRB, Sachdev2011}. Due to these features as well as its exact solvability, the 1D transverse field Ising model (TFIM) has served as a benchmark in understanding a plethora of other condensed matter systems. 

Consequently, efforts have been made over the last few decades in searching for experimental realizations of the TFIM. This has resulted in the finding of many candidate materials exhibiting Ising-like behavior \cite{dunn2012testing,coldea2010quantum,churchill2024transforming,chaloupka2024emergent,rado1970magnetoelectric}. But, careful studies of these systems have since revealed unexpected features that are inconsistent with a simple Ising description, suggesting the presence of additional interactions \cite{dunn2012testing,dollberg2024lihof4,churchill2024transforming,Woodland2023PRB,ellis1971magnetic}.
\\

A recent example is that of BaCo$_2$V$_2$O$_8$ (BCVO), an anti-ferromagnet (AF) consisting of nearly isolated cobalt chains with a four-fold screw structure \cite{Wichmann1986_BaCo2V2O8,He2005crystal,He2006APL,Niesen2013PRB,Ideta2013JKPS,ideta201251,shen2019magnetic}. At low temperature, the system has Néel order with the easy-axis aligned with the chain direction, which can be captured with an Ising-like XXZ model with moderate anisotropy \cite{He2006APL,Niesen2013PRB,
Kawasaki2011PRB,Ideta2013JKPS,Wang2018PRL,kimura2009low,faure2019tomonaga,halati2023repulsively}. But, in the presence of transverse magnetic field, neutron scattering experiments have shown the emergence of anomalous staggered magnetization that is perpendicular to both the applied field and the easy-axis, which persists above the critical point \cite{faure2018topological,zou20218,halati2023repulsively}. Furthermore, it was found that the transition point greatly depends on the direction of the applied in-plane field, jumping from \mbox{$\sim 10$T} when along the [100] direction to almost 40T when along the [110] direction \cite{He2006APL,kimura2013collapse}. Both of these facts are inconsistent with the simple XXZ description. \\

To account for these discrepancies, phenomenological additions to the model have been put forth, most notably a site-dependent $g$-tensor with large anisotropy \cite{kimura2013collapse,faure2018topological,zou20218,halati2023repulsively,cui2019quantum}. Such a term is typically attributed to distortions of the crystal structure. The magnetic ions, here Co$^{2+}$, are surrounded by oxygen anions forming an octahedral cage. In an undistorted system, this local environment would have perfect cubic symmetry. However, in the actual material, the anions have slight buckling which lowers the symmetry of the octahedron, allowing for an anisotropic $g$-tensor. Furthermore, due to the screw-chain structure, this buckling varies from site to site, thus giving rise to site-dependence.

On the other hand, it was recently shown that a microscopic theory gives rise to isotropic Heisenberg ($J$), as well as anisotropic Kitaev ($K$) and Gamma $(\Gamma)$ exchange interactions (known as the $JK\Gamma$ model) via the spin-orbit coupling (SOC). Among them, the Kitaev and Heisenberg combine to generate Ising behavior\cite{churchill2024transforming} while, crucially, the Kitaev and Gamma generate bond-dependent interactions even in the absence of the distortions \cite{bhullar2026}. On the other hand, the octahedral cages retain their cubic symmetry, thus giving a fully isotropic $g$-tensor with no site-dependence, in contrast to previous phenomenological descriptions. While this microscopic model can qualitatively explain the evolution of the magnetic structure under an applied transverse magnetic field\cite{bhullar2026}, it is not yet clear if it accounts for the large difference in critical fields between [100]- and [110]-directions that was reported in experiments.
\\

Here, we unify the microscopic Kitaev and phenomenological XXZ descriptions by extending the spin-orbit-coupled exchange model to include distortions of the CoO$_6$ octahedra. We show that the symmetry lowering induced by the distortions not only generates a site-dependent anisotropic $g$-tensor but also modifies the exchange Hamiltonian through a distortion-induced staggered interaction that plays an essential role in the magnetic response. 

The remainder of this paper is organized as follows. In Sec.~\ref{sec:Exchange}, we derive the exchange Hamiltonian for both the ideal and distorted crystal structures and analyze the role of the various exchange interactions in zero magnetic field. In Sec.~\ref{sec:g-tensor}, we determine the anisotropic site-dependent $g$-tensor generated by the octahedral distortions using a point-charge model. In Sec.~\ref{sec:field}, we investigate the magnetic-field response and show how the interplay between the exchange interactions and the $g$-tensor accounts for the large anisotropy of the critical field. Finally, Sec.~\ref{sec:conclusion} summarizes our conclusions.

\section{Generic Nearest-Neighbor 1D Spin Model}\label{sec:Exchange}

We first present the microscopic exchange Hamiltonian for screw-chain systems such as BaCo$_2$V$_2$O$_8$.
In the absence of the octahedral distortions, it reduces to the model proposed in \cite{bhullar2026}. The magnetic ions, Co$^{2+}$, have $3d^7$ electronic configuration. Hund's rules thus give total angular momenta $L_{\text{tot}}=3$, $S_{\text{tot}}=3/2$. The surrounding anions, which form an octahedral cage then generate a crystal electric field (CEF) with cubic symmetry. This splits the different orbitals and the lowest lying manifold will be 12-fold degenerate, with $L_{\text{eff}}=1$ and $S=3/2$. Adding SOC then further splits it into $J_{\text{eff}}$ multiplets, with the lowest energy being a $J_\text{eff}=1/2$ doublet \cite{churchill2024transforming, Liu2018PRB, Sano2018PRB, Liu2020PRL,Matsuda2025RMP,kim2023bond,halloran2023geometrical,van2023electronic,piwowarska2019origin}. From now on, we will denote this doublet at a site $i$ with the conventional notation $\boldsymbol{S}_i$ unless specified otherwise.

\begin{figure}
    \centering
    \includegraphics[width=0.99\linewidth]{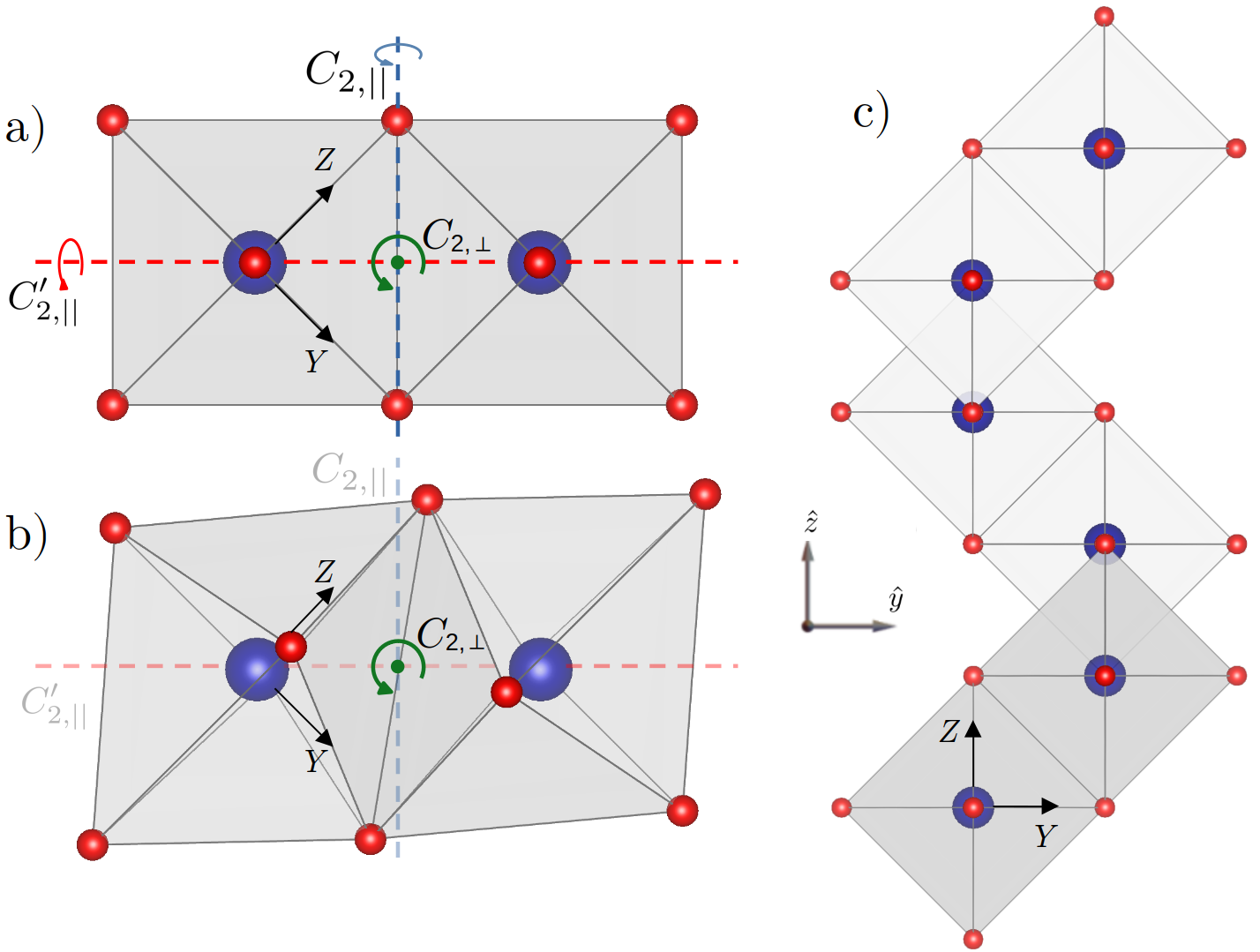}
   \caption{Geometry of a nearest-neighbor (n.n.) $X$-bond in the (a) ideal and (b) distorted structures. Note that the bond center is intersected by three mutually orthogonal two-fold rotation axes. The rotation axis $C_{2,\perp}$ (green) enforces $\Gamma'=\Gamma''=0$. In the ideal structure, in addition to $C_{2,\perp}$, the two additional two-fold rotation axes $C_{2,\parallel}$ and $C'_{2,\parallel}$ (blue and red) are present, which requires $K'=0$. In the distorted structure, the latter two symmetries are broken whereas $C_{2,\perp}$ is preserved, allowing only a finite $K'$. (c) Local $XYZ$ coordinate system associated with an $X$-bond (highlighted in gray) relative to the global $xyz$ axes. Successive bonds are related by a four-fold screw rotation about the $z$ axis, forming the screw chain.} 
    \label{fig:bond}
\end{figure}

Having established the atomic wave function, we consider the nearest neighbor (n.n.) exchange interaction in the local $XYZ$ coordinate system, whose axes point toward the corners of the CoO$_6$ octahedra. We begin by examining the geometry and symmetries of a representative bond. 
The two neighboring octahedra share an edge that lies in the plane defined by two of the local axes. For example, as illustrated in Fig.~\ref{fig:bond}(a), the shared edge lies in the $YZ$ plane, so that the $X$ axis is perpendicular to the edge-sharing plane. This bond is therefore referred to as an $X$-bond. Owing to inversion symmetry about the bond center, the most general n.n. exchange Hamiltonian contains six independent exchange parameters \cite{churchill2024transforming}  and can be written as
\begin{align}
   \label{eq:DistJKG}
    H_{ij}^\gamma &= J\boldsymbol{S}_i\cdot\boldsymbol{S}_j+KS_i^\gamma S_j^\gamma+\Gamma\left(S_i^{\alpha}S_j^\beta+S_i^\beta S_j^\alpha\right)\nonumber\\
   + & K^\prime S_i^\alpha S_j^\alpha + \Gamma^\prime\left(S_i^{\alpha}S_j^\gamma+S_i^\gamma S_j^\alpha\right)+\Gamma^{\prime\prime}\left(S_i^{\gamma}S_j^\beta+S_i^\beta S_j^\gamma\right),
\end{align}
where $\alpha$ and $\beta$ are the two local axes inside the edge-sharing plane, while $\gamma$ is normal to it. For an $X$ bond, this corresponds to $\gamma=X$, $\alpha=Y$, and $\beta=Z$.

As illustrated in Fig. \ref{fig:bond}, we can define three mutually orthogonal two-fold rotation axes denoted by $C_{2,\perp}$, $C_{2,\parallel}$, and $C_{2,\parallel}^\prime$ that intersect the bond center. Among them, the rotation axis $C_{2,\perp}$ (green) is a symmetry of the bond even in the distorted structure. Because this particular rotation sends
\begin{equation}
    (S^\alpha,S^\beta,S^\gamma) \rightarrow(-S^\alpha,-S^\beta,S^\gamma), 
\end{equation}
this symmetry enforces $\Gamma'=\Gamma''=0$. For this reason, we take these two parameters to be zero for the remainder of this work.

For the analysis that follows, it is useful to introduce global coordinates. We use the crystallographic $a$, $b$, and $c$ axes which we respectively denote as $x$, $y$, and $z$ coordinates.
As shown in Fig.~\ref{fig:bond}(c), the bonds in BCVO are arranged such that edge-sharing planes are normal to either the $x$ or $y$ axis, forming a screw-chain running along $z$ direction \cite{Wichmann1986_BaCo2V2O8,bhullar2026}. Consequently, there are bonds whose local $XYZ$ axes exactly coincide with the global $xyz$ coordinates, such as the $X$-bond highlighted in Fig.~\ref{fig:bond}(c).

Taking this as a starting point, the exchange Hamiltonian of each subsequent bond along the screw chain can then be obtained by repeatedly applying a four-fold screw-rotation along $z$. Therefore, in the global coordinates, the spin operators in Eq.~\ref{eq:DistJKG} will change from site to site such that $S_i^\gamma$ cycles through $S^x_i\to S_i^y\to -S^x_i\to-S_i^y\to S_i^x...$. After slight rearrangement of terms, the Hamiltonian for the entire chain is thus

\begin{eqnarray}
    \label{eq:fullJKG}
    &H_{xyz}&= \sum_{\langle ij \rangle} J\epsilon(S^x_iS^x_{j}+S^y_iS^y_{j})+JS^z_iS^z_{j}\nonumber\\
     &-(-1)^i&\frac{K-K^\prime}{2}(S^x_iS^x_{j}-S^y_iS^y_{j})\nonumber\\
     &+&\Gamma\cos\left(\frac{\pi i}{2}\right)(S^x_iS^z_{j}+S^z_iS^x_{j})\nonumber\\
    &+&\Gamma\sin\left(\frac{\pi i}{2}\right) (S^y_iS^z_{j}+S^z_iS^y_{j}),
\end{eqnarray}
where $\epsilon \equiv 1+\frac{K+K'}{2J}$ characterizes the XXZ anisotropy while $\frac{K-K'}{2}$ and $\Gamma$ parameterize the site-dependent exchange interactions. 

\subsection{Ideal Octahedral Case}

In the absence of distortions, each bond is symmetric under all three two-fold rotation axes, $C_{2,\perp}$ ,$C_{2,||}$, and $C_{2,||}^\prime$ (Fig.~\ref{fig:bond}(a)). These enforce $K^\prime=\Gamma^\prime=\Gamma^{\prime\prime}=0$. Consequently, Eq.~\ref{eq:DistJKG} reduces to the well-known $JK\Gamma$ model, and Eq.~\ref{eq:fullJKG} recovers the Hamiltonian proposed in Ref.~\cite{bhullar2026}. Since $K^\prime=0$, the XXZ anisotropy and the magnitude of the staggered interaction are $\epsilon=1+K/2J$ and $K/2$ respectively, implying that they are not independent but are both determined by the Kitaev interaction $K$. The resulting $JK\Gamma$ model captures the qualitative behaviors of BCVO including its sublattice magnetic structure and critical behaviours near the critical field.\cite{bhullar2026} 

An important question, though, is whether the  $JK\Gamma$ model based on the ideal crystal structure is sufficient to quantitatively account for experimentally observed features. To address this question, we first compare the spin dynamical structure factor (DSF), computed using 24-site exact diagonalization (ED), with zero-field inelastic neutron scattering (INS) measurements along the $(2,0,l)$ path in reciprocal space \cite{grenier2015longitudinal}. As shown in Fig.~\ref{fig:DSF} (left), the calculated spectrum shows a domain-wall excitation mode (highlighted in white) observed experimentally. However, its dispersion is systematically too small relative to the excitation gap, and we find no choice of $J$, $K$, and $\Gamma$ that quantitatively reproduces the INS spectrum.

\subsection{Distorted Octahedra Case}

To resolve the discrepancy discussed above, we now incorporate the effects of octahedral distortions. The microscopic arguments leading to Eq.~\ref{eq:DistJKG} remain valid, but the local cubic symmetry is lowered by the tilting and buckling of the anions. In BCVO, as illustrated in Fig.~\ref{fig:bond}(b), the $C_{2,\parallel}$ and $C_{2,\parallel}^\prime$ symmetries are broken, allowing a finite $K^\prime$. The $C_{2,\perp}$ symmetry, however, remains intact, enforcing $\Gamma^\prime=\Gamma^{\prime\prime}=0$.

Comparing with the ideal case, we note that the only modification is the appearance of a finite $K^\prime$ term. From Eq. \ref{eq:fullJKG}, this results in  the renormalization of the XXZ anisotropy, which becomes \mbox{$\epsilon\equiv 1+(K+K^\prime)/2J$}, as well as the coefficient of the staggered term, which becomes $(-1)^i (K-K^\prime)/2$. We see that these two quantities are decoupled when $K^\prime$ is finite. So, while the qualitative features of the ground state are not expected to change, this additional degree of freedom can allow for better quantitative agreement with experiments.\\

\begin{figure}[h!]
    \centering
    \includegraphics[width=1\linewidth]{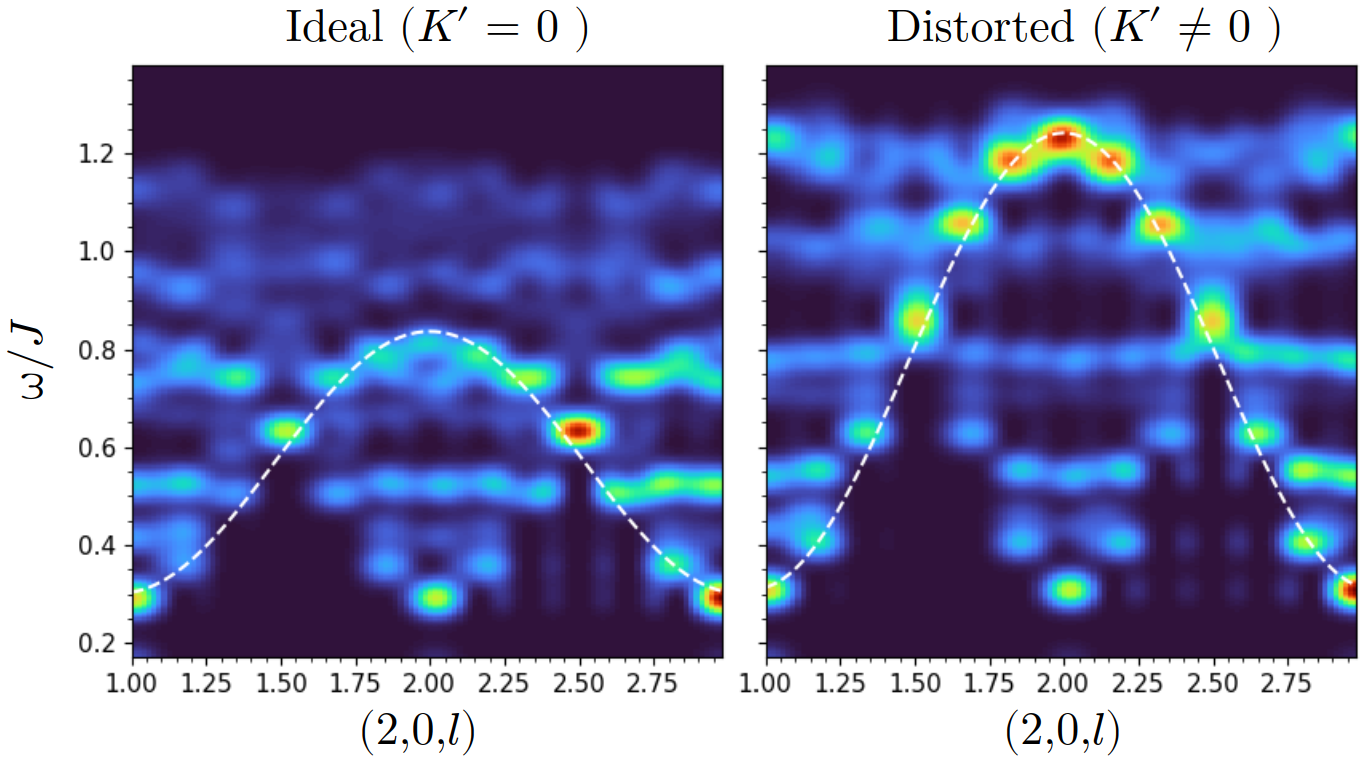}
    \caption{Dynamical structure factor (DSF) obtained by 24-sites ED with (a) $K'=0$ and (b) $K' \neq 0$ indicating the effects of distortion  via finite $K'$ interaction. In both, we use $J=5.8$meV, $\Gamma=0.6$ meV, and fix $\epsilon=0.56$. As $K^\prime-K$ becomes smaller, the domain-wall mode (white) becomes more dispersive while the gap remains nearly constant, consistent with the experimentally observed features\cite{grenier2015longitudinal} .}
    \label{fig:DSF}
\end{figure}

Indeed, repeating the ED calculation with $K^\prime\neq0$ (Fig.~\ref{fig:DSF}, right), we can obtain a more pronounced domain-wall mode with a larger dispersion. We find that agreement with INS spectrum \cite{grenier2015longitudinal} is achieved with $J=5.8$\,meV, a moderate XXZ anisotropy ($\epsilon=0.56$), and site-dependent interactions that remain small compared with $J$: $\Gamma=0.6$\,meV for the four-fold term, producing a small anticrossing at $l = n+1/2$\cite{takayoshi2023phase}, and $(K-K^\prime)/2=0.15$\,meV for the two-fold term. The latter requires $K^\prime\simeq K$ ($K=-2.7$\,meV and $K^\prime=-2.4$\,meV), demonstrating the crucial role of the distortion-induced interaction.\\

Physically, in the ideal structure a moderate XXZ anisotropy would require $|K|\sim J$, inevitably producing a staggered interaction of comparable magnitude. A finite $K^\prime$ lifts this constraint by independently controlling the XXZ anisotropy and the two-fold staggered interaction. As $K^\prime$ approaches $K$, the latter is strongly suppressed, increasing the dispersion of the domain-wall mode while leaving the excitation gap nearly unchanged. Consequently, the system is driven much closer to the XXZ limit, consistent with previous phenomenological model studies \cite{faure2018topological,grenier2015longitudinal,SKimura_2006,kawasaki2011magnetic,zhang2021quantum}.

Our microscopic analysis therefore leads to a distinct interpretation from the long-standing phenomenological picture \cite{faure2018topological,grenier2015longitudinal,SKimura_2006,kawasaki2011magnetic,zhang2021quantum,takayoshi2023phase}. It is often argued that BCVO is intrinsically described by an XXZ model and that octahedral distortions introduce weak site-dependent interactions. Instead, we find that the undistorted system naturally hosts strong bond-dependent interactions arising from the Kitaev exchange, while the primary role of the distortion is to \textit{attenuate} these interactions, driving the system toward the phenomenological XXZ limit \cite{konieczna2025understanding}.

\section{g-tensor Anisotropy}\label{sec:g-tensor}
 We now move to the effects of distortions on the $g$-tensor. Given their important role in the exchange Hamiltonian, as was argued above, we expect them to also play an important one here. We thus calculate this single-site $g$-tensor microscopically to quantify these effects. 
 
 To illustrate this approach, let's start from the ideal case and then add the distortions. As was described previously, Co$^{2+}$ have a 3$d^7$ configuration, with $L_\text{tot}=3$ and $S_\text{tot}=3/2$. Cubic CEF yields a low-lying $L_\text{eff}=1$, $S=3/2$ manifold which is split into $J_\text{eff}$ multiplets by spin-orbit coupling $\lambda_{SOC}$, with a $J_\text{eff}=1/2$ doublet having the lowest energy \cite{churchill2024transforming, Liu2018PRB, Sano2018PRB, Liu2020PRL,Matsuda2025RMP,kim2023bond,halloran2023geometrical,van2023electronic,piwowarska2019origin}. Adding a Zeeman field lifts the twofold degeneracy, and the effective $g$-tensor $\boldsymbol{g}_\text{eff}$ is found by projecting this pertubation onto the ground-state manifold:
\begin{equation}
    \label{eq:zeeman}
    H_\text{Zeeman}=\mu_B\left(\boldsymbol{L}_\text{eff}+2\boldsymbol{S}\right)\cdot\boldsymbol{B}\xrightarrow{\text{proj.}}\frac{1}{2}\mu_B\left(\boldsymbol{g}_\text{eff}\cdot\boldsymbol{B}\right)\cdot\boldsymbol{\sigma},
\end{equation}
where $\boldsymbol{\sigma}$ is the Pauli vector. The $g$-tensor components are given by
\begin{gather}
    \label{eq:tensorComps}
    g_{\text{eff},x\alpha}=2\text{Re}\left(\langle-|L_\alpha+2S_\alpha|+\rangle\right),\nonumber\\
    g_{\text{eff},y\alpha}=2\text{Im}\left(\langle-|L_\alpha+2S_\alpha|+\rangle\right),\\
    g_{\text{eff},z\alpha}=\langle+|L_\alpha+2S_\alpha|+\rangle-\langle-|L_\alpha+2S_\alpha|-\rangle\nonumber,
\end{gather}
where $\alpha=x,y,z$, and $|\pm\rangle$ are the two states in the ground-state manifold \cite{rinkevicius2008degenerate,gradl2018asymmetric,abragam1951theory}. The splitting between the two states is thus \cite{gradl2018asymmetric,abragam2012electron}
\begin{equation}
    \label{eq:zeemanSplit}
    \Delta E=\mu_B|\boldsymbol{g}_\text{eff}\cdot\boldsymbol{B}|=\mu_B\sqrt{\boldsymbol{B}^T\boldsymbol{g}_\text{eff}^T\cdot\boldsymbol{g}_\text{eff}\boldsymbol{B}}.
\end{equation}
Since the splitting depends only on the quantity \mbox{$\boldsymbol{g}_\text{eff}^T\cdot\boldsymbol{g}_\text{eff}$}, we can symmetrize our $g$-tensor by instead using \mbox{$\boldsymbol{g}\equiv \sqrt{\boldsymbol{g}_\text{eff}^T\cdot\boldsymbol{g}_\text{eff}}$} in Eq. \ref{eq:zeeman}. From now on this will be the $g$-tensor that we use. For the ideal octahedra case, then the states $|\pm\rangle$ are simply $|J_{z}=\pm\frac{1}{2}\rangle$, and we obtain an isotropic $g$-tensor with diagonal entries equal to the Landé $g$-factor $g_J=4\frac{1}{3}$, as expected \cite{Liu2020PRL,abragam1951theory}.

Let us now consider distortions of a single octahedral cage environment. These break the cubic symmetry, which further splits the $L_\text{eff}=1$ manifold \cite{liu2022PRB,abragam1951theory,Liu2020PRL}. Carefully examining the position of the surrounding anions, we find that in BCVO there still remains a $C_2$ axis around the $x\pm y$ direction, depending on which site we're considering \cite{Wichmann1986_BaCo2V2O8}. The most general terms that capture this~are 
\begin{multline}
    \label{eq:distHamiltonian}
    H_{\text{dist}}=\Delta_\tau\left(L^2_z-\frac{2}{3}\right)+\Delta_\omega\left(L_xL_y+L_yL_x\right)\\
    +\Delta_\mu\left(\left(L_x\mp L_y\right)L_z+L_z\left(L_x\mp L_y\right)\right),\\
\end{multline}
where the three $\Delta$'s are parameters to be determined. They modify the low-lying doublet with which we can re-calculate the $g$-tensor. Following the procedure outlined above, the resulting symmetrized $g$-tensor will have the general form
\begin{equation}
    \label{eq:genTensor}
    \boldsymbol{g}=\begin{pmatrix}
        g_{xx}&g_{xy}&g_{xz}\\
        g_{xy}&g_{xx}&g_{yz}\\
        g_{xz}&g_{yz}&g_{zz}\\
    \end{pmatrix},
\end{equation}
where the remaining $C_2$ symmetry has enforced $g_{xx}=g_{yy}$ and $|g_{xz}|=|g_{yz}|$.\\

Before moving on, let's examine the effect of each term in Eq. \ref{eq:distHamiltonian} on the resulting $g$-tensor. The first originates from tetragonal distortion along $z$. By itself, it generates only out-of-plane anisotropy (i.e. $g_{zz}\neq g_{xx}$, but $g_{xy}=g_{xz}=0$). The second one appears due to the orthorombic distortions of the octahedral cage and gives rise to in-plane anisotropy ($g_{xy}\neq 0$). Finally, the last term  allows for $g_{xz}\neq 0$. This introduces finite tilting of the principal axes away from the $z$-axis by some angle $\theta$.

\begin{figure}[h!]
    \hspace{-5mm}
    \includegraphics[width=1\linewidth]{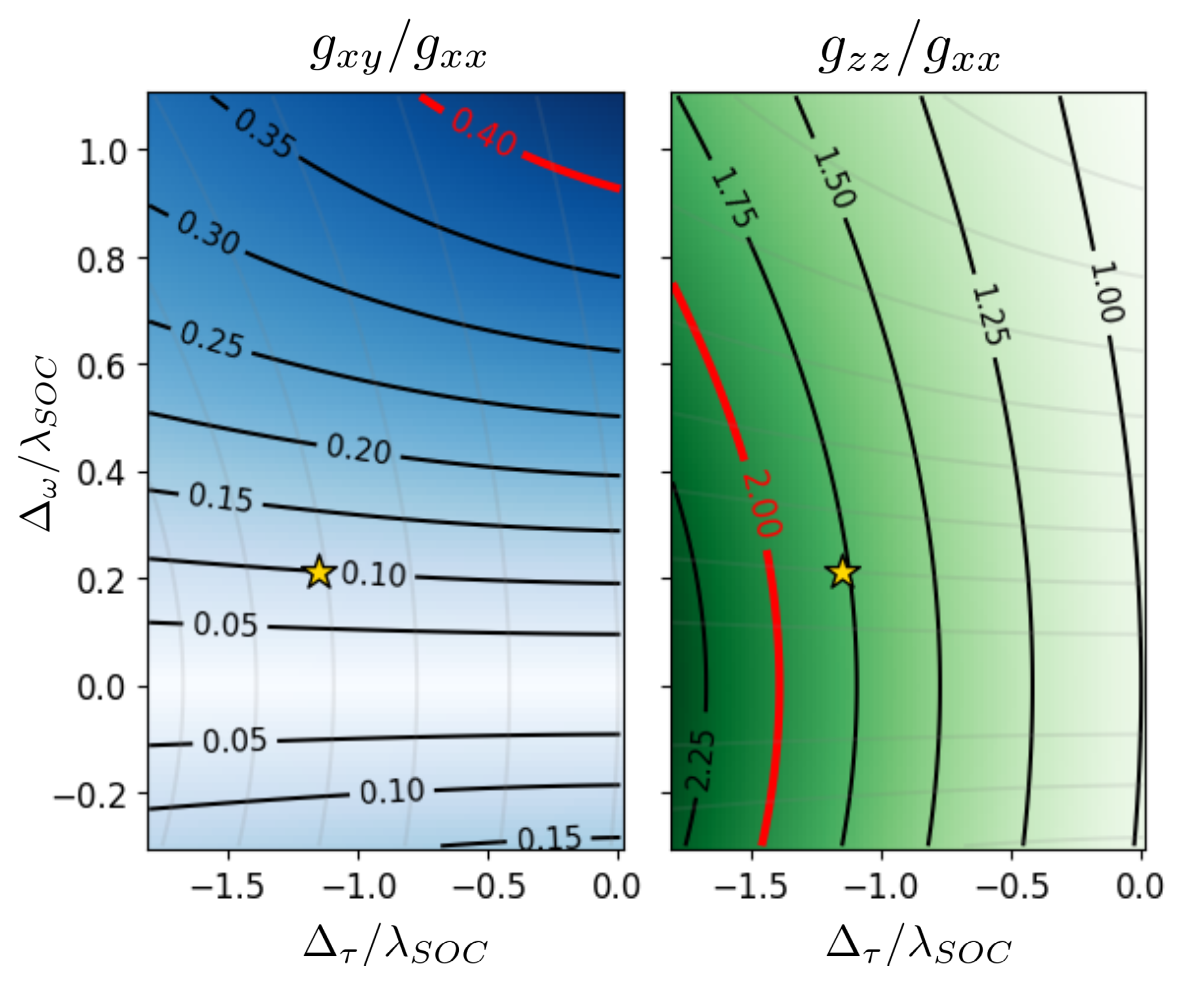}
    \caption{In-plane (left) and out-of-plane (right) anisotropies of $g$-tensor components for different distortion parameters ($\Delta_\mu/\lambda_{SOC}$ is fixed at -0.15). The red contour lines correspond to the value used in phenomenological descriptions ($g_{xy}/g_{xx}\approx0.4$ and $g_{zz}/g_{xx}\approx2$). We see that these cannot be obtained simultaneously for any given set of crystal field potentials $(\Delta_\tau,\Delta_\omega,\Delta_\mu)$. The star indicates the set of $\Delta$s obtained from our point charge calculation and the corresponding $g$-tensor components.}
    \label{fig:distParams}
\end{figure}

The dependence of the $g$-tensor anisotropy $g_{xy}/g_{xx}$ and $g_{zz}/g_{xx}$ on the crystal-field splittings $\Delta_\tau$ and $\Delta_\omega$ is shown in Fig.~\ref{fig:distParams}, with contour lines added. Previous phenomenological studies typically assumed a highly anisotropic $g$-tensor, with $g_{zz}/g_{xx}\approx2$ and $g_{xy}/g_{xx}\approx0.4$ \cite{faure2018topological,kimura2013collapse}, indicated by the red contours. However, these two lines do not intersect, indicating that there is no common set of crystal-field parameters $\Delta$s that can simultaneously reproduce both phenomenological $g$-factor ratios.

We thus estimate a set of splittings $\Delta_\tau$, $\Delta_\omega$ and $\Delta_\mu$ relevant to BCVO and calculate the resulting $g$-tensor values. To do so, we use a point charge approximation, starting from the full crystal field Hamiltonian
\begin{equation}
    \label{eq:cef}
    H_{CEF}=\sum_{n,m}B^m_nO_n^m,
\end{equation}
where $O^m_n$ ($n\leq4,|m|\leq n$) are the Stevens operators for $L_\text{tot}=3$, and find the coefficients $B^m_n$ from the electrostatic potential generated by the surrounding anions as well as other known atomic parameters \cite{stevens1952matrix,hutchings1964point,zvyagin2025stevens,kabeya2022eigenstate,dun2021effective}. We then project this Hamiltonian into the $L_\text{eff}=1$ subspace to recover the form of Eq. \ref{eq:distHamiltonian}. The coefficients are thus given by
\begin{align}
    \label{eq:deltas}
    \Delta_\tau &= 8B^0_2+20B^4_4-100B^0_4\\
    \Delta_\omega&=\frac{2}{3}B^{-2}_2+40B^{-2}_4\\
    \Delta_\mu &= -\frac{2}{3}B^{-1}_2+20B^{-1}_4-20B^{-3}_4
\end{align}
The different $B^m_n$ were calculated with the help of the PyCrystalField package \cite{scheie2021pycrystalfield}. Combining with standard literature value for SOC in Co$^{2+}$ \cite{abragam1951theory,Liu2020PRL,riedel2025applied}, this gives $(\Delta_\tau,\Delta_\omega,\Delta_\mu)/\lambda_{SOC}\sim(-1.20,0.20,-0.15)$.\\

These crystal-field parameters yield $g_{zz}=5.99$, \hbox{$g_{xx}=3.38$}, $g_{xy}=0.34$, and $g_{xz}=0.17$, as marked by the star in Fig.~\ref{fig:distParams}. This corresponds to $g_{zz}/g_{xx}\approx1.77$ and $g_{xy}/g_{xx}\approx0.10$, and a tilting angle of $\theta\sim 4.5^\circ$, providing a microscopic benchmark for the $g$-tensor anisotropy in BCVO.

Once we successively apply a $4_1$ screw rotation along $z$, we obtain the $g$-tensor at the remaining sites in the chain. Our theory then recovers the site-dependence of the components used in phenomenological models \cite{faure2018topological,kimura2013collapse}: $g_{xy}\to(-1)^jg_{xy}$, \hbox{$g_{xz}\to\sqrt{2}\sin(\frac{\pi}{4}+\frac{\pi j}{2})g_{xz}$, and $g_{yz}\to\sqrt{2}\cos(\frac{\pi}{4}+\frac{\pi j}{2})g_{yz}$}.\\

\section{Phase Transition Under External Field: Critical Field Anisotropy}\label{sec:field}

\begin{figure}[h!]
    \centering
    \includegraphics[width=0.85\linewidth]{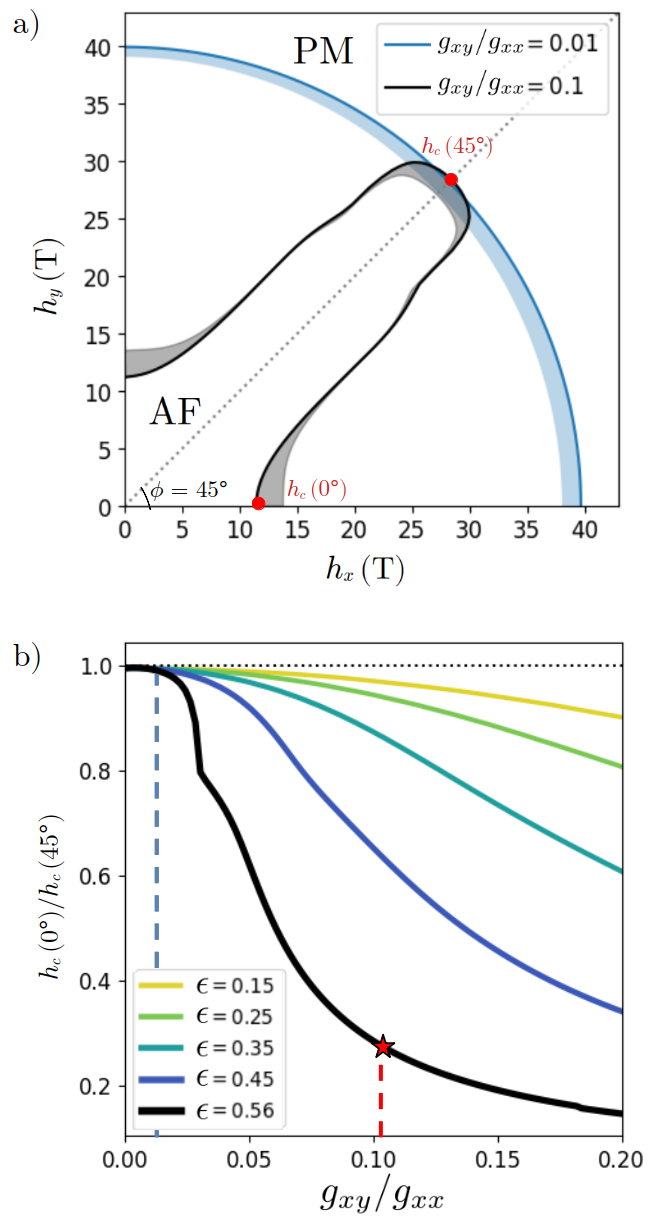}
    \captionsetup{skip=2pt}
    \caption{(a) Phase boundary $h_c(\phi)$ for applied transverse field [$h_x,h_y,0$] with fixed $\epsilon=0.56$. Using the calculated $g$-tensor $g_{xy}/g_{xx}$ (black), we see a large in-plane anisotropy with $h_c(0^\circ)\approx11$T and $h_c(45^\circ)\approx38$T indicated by the red dots. This anisotropy nearly vanishes when reducing $g_{xy}/g_{xx}$ (blue). For fixed $g_{xy}/g_{xx}$, varying the site-dependent exchange (here $|K-K^\prime|\leq1$meV and $|\Gamma|\leq2$meV) only slightly alters the respective phase boundary which remains roughly within the shaded region. (b) Anisotropy between $\phi=0^\circ$ and $\phi=45^\circ$ critical fields as a function of $g_{xy}/g_{xx}$ for various values of $\epsilon$. As $\epsilon$ gets smaller, their ratio rapidly approaches 1. The star indicates parameters calculated microscopically.}
    \label{fig:CritFields}
\end{figure}

With the exchange Hamiltonian established and the microscopic $g$-tensor determined, we now combine the two to investigate the in-plane (perpendicular to chain direction) anisotropy of the critical field. For a given in-plane field direction $\phi$, we calculate the ground state as a function of the field magnitude using DMRG \cite{White1992PRL, schollwock2011density,fishman2022itensor} and identify the transition to paramagnetic (PM) state from the peak in the staggered magnetic susceptibility \cite{bhullar2026,bhullar2025field,Sorensen2023PRRl}. Using the parameters obtained above, we find the phase boundary $h_c(\phi)$ shown in \hbox{Fig.~\ref{fig:CritFields}(a) (black).} The critical fields are $h_c(0^\circ)\approx11$\,T for the [100] direction and $h_c(45^\circ)\approx38$\,T for the [110] direction, which is consistent with experiments \cite{He2006APL,kimura2013collapse}.

Since both the exchange Hamiltonian and the $g$-tensor break rotational symmetry about the z axis, each can contribute to the observed in-plane anisotropy. To identify their respective roles, we vary the relevant parameters independently and examine the resulting changes in the phase boundary.

We begin by varying the size of the site-dependent exchange interactions ($K-K^\prime$ and $\Gamma$ in Eq. \ref{eq:fullJKG}).
We find that neither significantly affects the phase boundary, which roughly remains within the shaded region of \hbox{Fig. \ref{fig:CritFields}(a).} This indicates that the Kitaev-like site-dependent interactions alone cannot account for the large in-plane critical-field anisotropy observed in BCVO.

We next examine the individual components of the $g$-tensor. Varying the out-of-plane component $g_{zz}$ or the tilting angle $\theta$ (controlled by $g_{xz}$) has little effect on the phase boundary. In contrast, even small changes in $g_{xy}$ strongly modify the critical field anisotropy. As $g_{xy}$ approaches zero, the critical fields become nearly identical for all in-plane field directions (Fig. \ref{fig:CritFields}(a), blue), identifying $g_{xy}$ as the primary origin of the anisotropy.

While $g_{xy}$ is a primary drivers of the in-plane critical field anisotropy, the magnitude of its effect is strongly influenced by the exchange anisotropy $\epsilon$. For a fixed value of $g_{xy}$, reducing $\epsilon$ strongly suppresses the critical field anisotropy, as shown in Fig.~\ref{fig:CritFields}(b).\\

The interplay between $g_{xy}$ and $\epsilon$ can be understood from a simple classical picture. For simplicity, we neglect $g_{xz}$ and $g_{yz}$. An applied magnetic field favors a polarized state, competing with the antiferromagnetic order stabilized by the exchange interaction ($J>0$). Above a critical field, the polarized state becomes energetically favorable, giving rise to the phase transition. In the presence of off-diagonal components of the $g$-tensor, however, a magnetic field $\boldsymbol{h}$ generates an effective field
\begin{equation}
\boldsymbol{h}_{\rm eff}=\boldsymbol{g}\cdot\boldsymbol{h},
\end{equation}
whose direction depends on the orientation of the applied field. Consequently, the spins in the polarized phase align with $\boldsymbol{h}_{\rm eff}$ rather than the applied field itself.

For example, for the field along [110], $\boldsymbol{h}=h\left[\boldsymbol{\hat{x}}+\boldsymbol{\hat{y}}\right]$. The resulting field felt by a spin on site $j$ thus has the form 
\begin{equation}
    \label{eq:h45}
    \boldsymbol{h}_\text{eff}^{45^\circ}=(g_{xx}+(-1)^jg_{xy})\cdot h\boldsymbol{}\left[\boldsymbol{\hat{x}}+\boldsymbol{\hat{y}}\right].
\end{equation}

Thus, the effective field still points along [110], with only the magnitude at each site alternating between \hbox{$g_{xx}\pm g_{xy}$}. If $g_{xy}$ is not too large, then the critical field is not expected to be too different from the $g_{xy}=0$ case. On the other hand, for the field along [100], i.e. $\boldsymbol{h}=h\boldsymbol{\hat{x}}$ the effective field felt by each spin is now
\begin{equation}
    \label{eq:h0}
    \boldsymbol{h}_\text{eff}^{0^\circ}=h\left[g_{xx}\boldsymbol{\hat{x}}+(-1)^jg_{xy}\boldsymbol{\hat{y}}\right],
\end{equation}
where the magnitude is uniform for all sites, but more importantly there is now an additional staggered component perpendicular to the applied field. 
The polarized state therefore retains a staggered in-plane component whose magnitude is set by $g_{xy}$. Because $J>0$ and $\epsilon\neq0$, this staggered component is partially compatible with the antiferromagnetic exchange and therefore reduces the exchange energy cost of the polarized state. As a result, the transition occurs at a lower critical field. Reducing $\epsilon$ weakens this energetic gain, so the critical field increases and the in-plane anisotropy is correspondingly suppressed.

\section{Summary and Discussion}\label{sec:conclusion}
In summary, we have developed a microscopic description of BCVO that unifies its exchange interactions and $g$-tensor within a common framework that incorporates the local crystal-field environment. Starting from the distorted CoO$_6$ octahedra, we derived the most general n. n. exchange Hamiltonian and demonstrated that structural distortions generate an additional Kitaev interaction $K^\prime$. This interaction decouples the XXZ anisotropy from the site-dependent exchange interactions, providing a quantitative description of the zero-field spin dynamics \cite{grenier2015longitudinal} while naturally explaining why BCVO appears phenomenologically XXZ-like \cite{faure2018topological,grenier2015longitudinal,SKimura_2006,kawasaki2011magnetic,zhang2021quantum} despite its underlying Kitaev origin.

Our microscopic crystal-field analysis further shows that the $g$-tensor is considerably less anisotropic than assumed in previous phenomenological models \cite{faure2018topological,kimura2013collapse}.  Nevertheless, the off-diagonal component $g_{xy}$ plays a crucial role in determining the field response. By combining the microscopic exchange Hamiltonian with the calculated $g$-tensor, we reproduce the experimentally observed in-plane anisotropy of the critical field \cite{He2006APL,kimura2013collapse}. A systematic analysis reveals that the bond-dependent exchange interactions contribute only weakly to this anisotropy, whereas a combination of $g_{xy}$ and XXZ anisotropy $\epsilon$ is its primary origin; the exchange anisotropy ($\epsilon$) controls the effectiveness of $g_{xy}$ by determining the exchange-energy gain associated with the staggered component of the polarized state. The observed critical-field anisotropy therefore emerges from the cooperative interplay between the microscopic exchange interactions and the crystal-field-induced $g$-tensor, rather than from either mechanism alone. More broadly, our work establishes a microscopic framework for understanding 1D Ising materials beyond BCVO through exchange interactions and the $g$-tensor.

\section*{Acknowledgments} This work is supported by the NSERC Discovery Grant No. 2022-04601. H. Y. K. acknowledges support from the Canada Research Chairs Program No. CRC-2019-00147.
This research was enabled in part by support provided by Compute Ontario,  Calcul Québec, and the Digital Research Alliance of Canada.

\bibliography{references} 
\newpage
\end{document}